\documentclass[twocolumn,showpacs,amsmath,amssymb,floatfix,prl]{revtex4}

\usepackage{graphicx}
\begin{document}

\title{Ballistic Thermal Conduction across Acoustically Mismatched
Solid Junctions}

\author{Jian Wang}
\author{Jian-Sheng Wang }

\affiliation{Department of Computational Science,
National University of Singapore, Singapore 117543, Republic of Singapore}

\date{17 May 2005}

\begin{abstract}
We derive expressions for energy flow in terms of lattice normal mode
coordinates and energy transmission involving reduced group
velocities. With a version of Landauer formula appropriate for lattice
dynamic approach, the phonon transmission coefficients and thermal
conductance are calculated for two kinds of acoustically mismatched
junctions: different chirality nanotubes $(11,0)$ to $(8,0)$, and
Si-Ge superlattice structure.  Our calculation shows a mode-dependent
transmission in nanotube junction and a resonantly modulated ballistic
thermal conductance in superlattice.  The superlattice result suggests
a new interpretation of the experimental data.  Our approach provides
an atomistic way for the calculation of thermal conduction in
nanostructure.
\end{abstract}


\maketitle

Rapid progress in the synthesis and processing of materials with
structures of nanometer length scales has created a demand for
the understanding of thermal transport in nano-scale low dimensional
devices \cite{Baowen,Jswang,K.Schwab,dgcahill}. Nanostructures offer a
new way of controlling thermal transport by tuning dispersion
relations and other parameters \cite{Baowen}. Recent 
experimental and theoretical studies have revealed novel features of phonon
transport in these systems, such as the size-dependent anomalous heat
conduction in one-dimensional (1D) chain \cite{Jswang} and the
universal quantum thermal conductance \cite{K.Schwab}.  Thermal
transport in nanostructures may differ from the predictions of Fourier's
law based on bulk materials; this may happen because of the existence
of many acoustically mismatched interfaces in nanostructures and
because the phonon mean free path is comparable to the size of the
structure \cite{dgcahill}.  An understanding of the thermal conduction
across acoustically mismatched solid interfaces is a necessary
requirement for thermal transport engineering.

The study of thermal transport across interfaces dates back as early
as to 1940s when Kapitza resistance \cite{ETswartz} was reported and
much work has been done in this field \cite{dgcahill,ETswartz}.  In
general, theoretical modeling of this problem has been undertaken
either by the acoustic-mismatch model (AM) with scalar elastic waves,
or by the diffuse-mismatch model (DM) with Boltzmann transport
equation \cite{G-Chen}. Some numerical methods such as molecular
dynamic simulation \cite{MD} have also been used.  While AM and DM
models provide some useful reference calculations, scalar wave model
and Boltzmann transport equation are only phenomenological
descriptions and they have ignored the complexity of the
interface. Atomic-level lattice dynamic (LD) approach should be the
right way of capturing the mechanism of heat transport.  However,
after its early proposal in Ref.~\cite{dayang}, there has been little
further work using this approach.

Landauer formula for ballistic heat transport has been used
\cite{gcrego} for the prediction of universal quantum heat conductance
at very low temperatures.  The formula derived under continuum
assumption cannot be applied straightforwardly to systems on nanoscale
where atomic details are important.  In this paper,
we outline a new derivation from the point of view of lattice
dynamics.  The lattice formulation takes into account the different masses
of the atoms and various vibrational modes.  We also propose a method
of computing the transmission coefficients by solving a set of
dynamical equations with scattering boundary conditions.  The method
is applied to compute the transmission coefficients of a
carbon-nanotube junction and superlattice structures, and the thermal
conductance is obtained with Landauer formula.

We consider systems with perfect leads on the left and right with an
arbitrary interaction at the junction.  The Hamiltonian for such a
system of vibrating atoms under linear approximation is given by
\begin{equation}
  \label{hamilton}
H = \sum\limits_l \Bigl( \sum\limits_{i,\alpha} \frac{{p_{l,i}^\alpha} ^2}{2m_i} +
\!\!\!\!\sum\limits_{l'\!,i,\alpha ;j,\beta }\!\! \frac{1}{2}
K_{l,i;l'\!\!,j}^{\alpha,\beta} u_{l,i}^\alpha u_{l'\!\!,j}^\beta  \Bigr)
 = \sum\limits_l \varepsilon_l,
\end{equation}
where $l$ or $l'$ denotes a unit cell, $i,j$ the position in a cell,
$\alpha, \beta$ the direction of vibrating motions of atoms, and
$u_{l,i}^\alpha$ the displacement from equilibrium of the atom $(l,i)$
with equilibrium position ${\bf R}_{l,i}$.  A local energy density can
be defined through the energy in cell $l$, as $\rho({\bf r}) =
\sum_{l} \varepsilon_l \delta({\bf r} - {\bf R}_l)$ where ${\bf R}_l$
is a lattice vector.  An expression for the heat current in the $z$
direction can be derived from the energy continuity equation,
$\partial \rho/\partial t + \nabla\! \cdot {\bf j} = 0$, as
\begin{equation}
  \label{energyflow}
  \bar {I}_z = \frac{1}{L_z} \lim_{q_z \to 0}
\frac{ \left\langle {\dot {\Omega }
({\bf q} ,t)} \right\rangle _t }{ - iq_z }.
\end{equation}
In this equation, $q_z $, $\Omega ({\bf q}, t)$ and $L_z$ are the
z-component momentum, the energy density in momentum space and the
system length along $z$ direction, respectively, with time average
represented by $ \left\langle \ \right\rangle_t $.  The energy density
in momentum space is given by
\begin{eqnarray}
 \label{energydensitynormal}
\Omega \left({\bf q}_0 ,t\right) =
\sum_l \varepsilon_l e^{ - i {\bf q}_0 \cdot {\bf R}_l }
=\frac{1}{2}\!\!\!\!\sum_{{\bf q}, {\bf q}', n, n'}\!\!\! \Big\{ \big(
\dot{Q}_{{\bf q}}^n \dot{Q}_{{\bf q}'}^{n'} \nonumber\\
  +\;\omega_{n'}^{2}({\bf q}')
{Q}_{{\bf q}}^n {Q}_{{\bf q}'}^{n'} \big)
 \times \sum_{i,\alpha} \tilde
{e}_{i,n}^\alpha({\bf q}) \tilde {e}_{i,n'}^{*\alpha}({\bf q}')
\delta_{{\bf q} + {\bf q}', {\bf q}_0}\Big\},
\end{eqnarray}
where $Q_{\bf q}^n$'s are normal mode coordinates for the $n$-th
branch phonons, with mode $\tilde {u}_{l,i,n}^\alpha (t) =
\frac{1}{\sqrt {m_i } }\tilde {e}_{i,n}^\alpha ( {\bf q}) e^{i({\bf q}
\cdot {\bf R}_l - \omega t)}$, where
$\sum_{i,\alpha}\tilde{e}_{i,n}^\alpha({\bf q})
\tilde{e}_{i,n'}^{*\alpha}({\bf q}) =\delta_{n,n'}$
\cite{LDtextbook}. Combining Eq.~(\ref{energyflow}) and
Eq.~(\ref{energydensitynormal}), the time-averaged energy current
along z-direction is
\begin{equation}
  \label{energyflowq}
  \bar {I}_z =\frac{-i}{L_z}\sum\limits_{{\bf q},n}
{\omega_{n}({\bf q}) \frac{\partial \omega_{n}({\bf q})}{\partial q_z} \langle
Q_{\bf q}^{n} \dot{Q}_{\bf q}^{n*}  \rangle_t}.
\end{equation}
The transmission coefficients are defined with respect to the
normalizations of incoming and outgoing waves.  The total energy
current from one particular lead, say the left lead, is an arbitrary
superposition of all the modes.  Thus, the motion of the atoms is
described by the wave:
\begin{eqnarray}
  \label{wave}
  \Psi _{l,i}^{\alpha ,L} \left( {\omega ,t} \right) &=& \sum_n
a_n^L \tilde {u}_{l,i,n}^{\alpha ,L}(\omega, {\bf q})  \nonumber \\
&+&\!\! \sum_n \Bigl(
\!\!\sum_{n',\sigma = L,R} \!\!
a_{n'}^\sigma t_{n\,n'}^{L\,\sigma}  \Bigr)
\tilde {u}_{l,i,n}^{\alpha ,L}(\omega , -{\bf q}'),
\end{eqnarray}
where $a_n^\sigma$ is the amplitude of the mode $n$ in lead $\sigma$,
while $t_{n\,n'}^{\sigma\,\sigma'}$ is the transmission/reflection
amplitude from mode $n'$ in lead $\sigma'$ to mode $n$ in lead
$\sigma$.  Note that $\bf q$ and ${\bf q}'$ satisfies
$\omega=\omega_n({\bf q}) = \omega_n({\bf q}')$.  Similar expression
can be written down for the right lead.  Since the energy is
conserved, there is no net energy accumulation in the junction, which
means that the time-averaged energy currents from both sides are equal,
$\bar {I}^L \equiv \bar {I}^R$.  This condition leads to the following
identity for the transmission amplitudes:
\begin{equation}
  \label{transmission}
  \sum\limits_{\sigma ,n}{ t_{nn^{\prime}}^{\sigma\sigma^{\prime}}
  t_{nn^{\prime\prime}}^{*\sigma\sigma^{\prime\prime}}\tilde
  {v}_n^\sigma}
  = \tilde {v}_{n^{\prime}}^{\sigma^{\prime} } \delta _{n^{\prime}\sigma
^{\prime},n^{\prime\prime}\sigma ^{\prime\prime}},
\quad \tilde {v}_n^\sigma = {v_n^\sigma }/{\emph{l}_{z}^\sigma }.
\end{equation}
The important difference with the continuum case \cite{gcrego} is that
we need to replace the group velocity with a reduced group velocity
$\tilde {v}_n^\sigma = {v_n^\sigma }/{\emph{l}_{z}^\sigma }$, where
$\emph{l}_{z}^\sigma$ is the length of unit cell along the $z$
direction, $v_n^\sigma = \partial \omega^\sigma_n/\partial q_z$.  For 1D
quantum thermal energy current, with the help of the
relation~(\ref{transmission}), after quantizing
Eq.~(\ref{energyflowq}), we get the formula
\begin{equation}
  \label{landuerequation}
\bar {I} = \frac{1}{2\pi }\sum\limits_n {\int_{\omega _{\min }
}^{\omega _{\max } }\!\!\!\!\! d \omega\;{\hbar \omega \Bigl(f (\omega ,T_L ) -
f (\omega ,T_R )\Bigr)T_n (\omega )} },
\end{equation}
where $T_{n}\left( \omega \right) =\sum\limits_{m}{\left\vert
{t_{mn}^{RL}}\right\vert ^{2}{\tilde{v}_{n}^{R}}/{\tilde{v}_{m}^{L}}}$
is the energy transmission probability, $f(\omega, T_\sigma)$ is
the Bose-Einstein distribution for the left or right lead.
Eq.~(\ref{landuerequation}) is the Landauer formula for quantum
thermal energy flow.  The thermal conductance is obtained by taking an
infinitesimal temperature difference.

The central issue now is to have an efficient method to calculate the
transmission coefficients across junctions at the atomistic
level. This appears to be a difficult problem \cite{dayang} in general
taking into account the complexity of the interface.  Transfer matrix
method can be used for simple 1D models with a few interfaces.  But
this method could not be applied to a large 1D system or 3D
systems due to numerical instability.

We propose what we call the scattering boundary equation (SBE) method
as a numerically more satisfactory solution.  Each atom in the system
satisfies the dynamic equation, $-m_i \omega ^2 {\bf u}_i +
\sum\nolimits_j K_{ij} ( {\bf u}_i - {\bf u}_j ) = 0 $, where the $3\times
3$ matrix $K_{ij}$ is the force constants between atom $i$ and
$j$. The boundary conditions are of the form $ {\bf \tilde u }_{q,n'}
+ \sum_n r_n { \bf \tilde u}_{-q,n}$ for the incoming waves and
$\sum_{n} t_n { \bf \tilde u'}_{q',n}$ for the outgoing waves, where
${ \bf \tilde u}_{q,n}$ and ${\bf \tilde u'}_{q',n}$ are the eigen
modes on the left and right leads, while the reflection and
transmission coefficients $r_n$ and $t_n$ are unknown. These equations
in matrix form are illustrated as
\begin{equation}
  \label{sbe}
\left( {{\begin{array}{*{10}c}
 \hfill & \cdots \hfill  & \hfill & \hfill \\
 \vdots \hfill & { K_{ii} } \hfill & { -
K_{ij} } \hfill & \hfill & \hfill \\
 \hfill  & \hfill & \hfill & \hfill \\
 1 \hfill & &  & { - \tilde u_{ - q,n} } \hfill & \hfill \\
 \hfill & \hfill & 1 \hfill & \hfill & { - \tilde u'_{ q',n} } \hfill \\
\end{array} }} \right)\left( {{\begin{array}{*{10}c}
 { u_{1\!\!\!\!\!} } \hfill \\
 \vdots \hfill \\
 {u_{m\!\!\!\!\!} } \hfill \\
 r_{n} \hfill \\
 t_{n} \hfill \\
\end{array} }} \right) = \left( {{\begin{array}{*{10}c}
 0 \hfill \\
 \vdots \hfill \\
 0 \hfill \\
 { \tilde u_{q,n\!\!\!\!\!} } \hfill \\
 0 \hfill \\
\end{array} }} \right)
\end{equation}
where $K_{ii}=-m_i \omega^{2}+\sum\nolimits_j {K_{ij} }$.  In 1D, the
matrix is square and can be both analytically and numerically solved
by the conventional method.  For higher dimensions, however, the
boundary conditions are complicated and the number of equations may be
larger than that of variables. But these equations are not linearly
independent and can still be solved by the singular value
decomposition method.

\begin{figure}[tb]
\includegraphics[width=0.9\columnwidth]{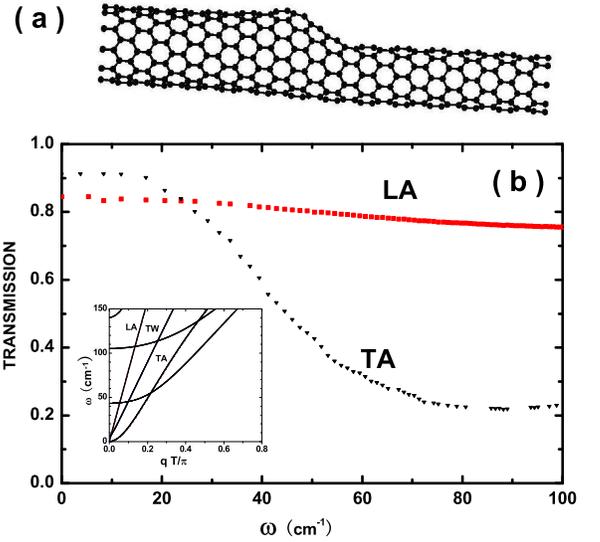}
\caption{\label{fig:nanotube}(a) Structure of a (11,0) and (8,0)
nanotube junction. (b) The energy transition $T_n(\omega)$ as a
function of angular frequency $\omega$.  The inset shows the
dispersion relation of the infinite-long (11,0) nanotube.}
\end{figure}

We first discuss the SBE results of the transmission coefficients for
a nanotube junction, shown in Fig.~\ref{fig:nanotube}. This
calculation involves $ 504 $ equations and $455$ variables.  The
semiconductor nanotube junction structure is optimized by a
second-generation Brenner potential \cite{dwbrenner} with the force
constants derived from the same potential. The phonon dispersion for
nanotube $\mbox{(11,0)}$ is illustrated in the inset, in which four
acoustic branches are considered: the longitudinal mode (LA), doubly
degenerate transverse mode (TA), and the unique twist mode (TW) in
nanotubes.  Although all modes of a given frequency are considered, we
did not find mode-mixing behavior among acoustic modes at the lower
frequency range. The transmission for LA mode stays around 0.8 with
only small changes.  This value is below the AM model prediction of $0.98$
with the longitudinal group velocity $20.18\; {\rm km}\!  \cdot\! {\rm
s}^{-1}$ and $20.95\; {\rm km}\! \cdot\! {\rm s}^{-1}$ for
$\mbox{(11,0)}$ and $\mbox{(8,0)}$. In contrast, the transmission for
TA mode decreases with frequency. This can be accounted for by a
nearly quadratic dispersion relation of the TA mode, as illustrated in
the inset.  The transmissions of the TW mode and many other optical
modes are nearly zero or very small. This appears related to the
difference in rotational symmetries of these modes.  We propose that
this kind of mode-dependent transmission behavior may be important for
further application such as phonon filters.

Next, we consider two acoustically mismatched chains of the same
atomic mass connected with springs of different stiffnesses $K_1, K_2$
and lattice constants $a, b $ on each side.  The energy transmission
by transfer matrix calculation is $\left| t
\right|^2\frac{\widetilde{v}_R }{\widetilde{v}_L } = {4\left(
{\frac{\widetilde{v}_R }{\widetilde{v}_L }} \right)^2}/{\left( {1 +
\frac{\widetilde{v}_R }{\widetilde{v}_L }} \right)^2}$ where
$\widetilde{v}_L = v_L / a$, $\widetilde{v}_R = v_R / b$. Using mass
densities $\rho _L = m / a$, $\rho _R = m / b$ and $z_L = \rho _L v_L
$, $z_R = \rho _R v_R $, the AM model gives \cite{walittle}
$\frac{4z_L z_R }{(z_L + z_R )^2} = {4\left( {\frac{\widetilde{v}_R
}{\widetilde{v}_L }} \right)^2}/{\left( {1 + \frac{\widetilde{v}_R
}{\widetilde{v}_L }} \right)^2}$, in agreement with our result. Thus,
for simple lattice structure and smooth junction, the results of LD
and AM are the same.

\begin{figure}[tb]
\includegraphics[width=0.9\columnwidth]{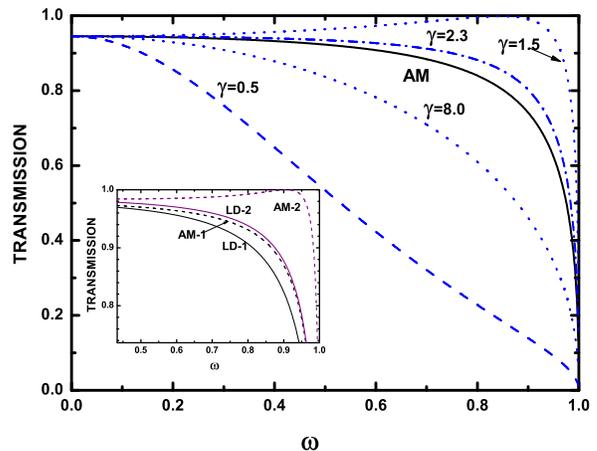}
\caption{\label{fig:miss}Energy transmission $T(\omega)$ as a function
of scaled angular frequency $\omega/\sqrt{4K_1}$.  Different curves
correspond to different coupling strength ratios $\gamma = K_f/K_1$ at
the junction with the spring stiffness parameters $K_2=2.6K_1$.  The
inset shows the difference in the prediction of energy transmission
for the composite lattice structure, $K_2=0.3846K_1$, $K_3=K_1$ for
AM-1 and LD-1; $K_2=1.1538K_1$, $K_3=3.0 K_1$ for AM-2 and LD-2.}
\end{figure}

For a more realistic model, not only the physical lattice may be
composite, but also the boundaries can be more complex.  We choose
effective stiffness to fit the diamond structure Silicon (Si) and
Germanium (Ge) phonon dispersion relations.  For Si and Ge along the
direction $\Gamma$-$X$, the highest frequency for acoustic phonon
branch is about $400\; {\rm cm}^{ - 1}$ and $250\;{\rm cm}^{-1}$,
respectively \cite{giannozzi}. If the phonon dispersion relation along
this direction can be regarded as one-dimensional \cite{phyldgaard},
the spring stiffness for each can be calculated by $\omega =
\sqrt{4k/m} = \sqrt {4K} $.  We use these experimental parameters
$K_{2}=2.6K_{1}$ for Si and Ge in the following discussion.

Fig.~\ref{fig:miss} shows how the energy transmission changes with the
boundary spring stiffness $K_f$.  It is reasonable to assume that the
stiffness $K_f$ for the Si-Ge junction is between pure Si and Ge
lattices.  When $\gamma=K_f/K_1=1.5$ or 2.3, the transmission is
higher than the prediction given by direct acoustic mismatch
model. Moreover, when the lattice structure is not a simple lattice,
we also find that AM model does not hold any more.  The inset in
Fig.~\ref{fig:miss} shows the discrepancy in the prediction of the
energy transmission by AM and LD model for a simple lattice with
stiffness $K_1 $ on the left and alternating stiffness $K_2$ and $K_3$
on the right.  Comparing with LD result, the prediction of AM model
overestimates the energy transmission at high frequency for mismatched
composite lattice structure.

\begin{figure}[t]
\includegraphics[width=0.90\columnwidth]{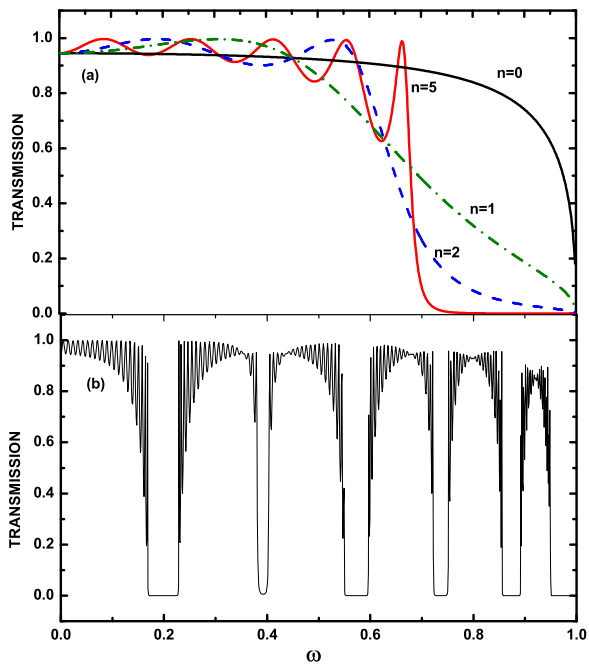}
\caption{\label{fig:modulation} Energy transmission $T(\omega)$ of a
superlattice with the Si and Ge leads; $\omega$ is scaled by
$\sqrt{4K_{1}} $. (a) Superlattice of alternating Si and Ge monolayer,
$n$ is the number of periods. (b) Superlattice of ten period layers,
each layer with $ 5\times 5 $ Si and Ge monolayers.}
\end{figure}
\begin{figure}[t]
\includegraphics[width=0.9\columnwidth]{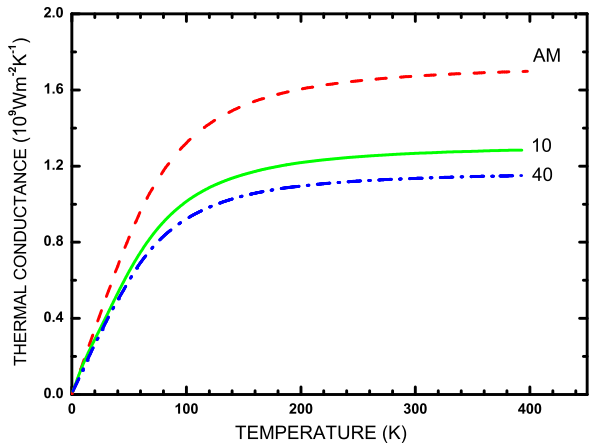}
\caption{\label{fig:exp}Thermal conductance as a function of
temperature. Curves labeled $10$ and $40$ represent ten superlattice
period layers, each layer with $5\times5$ and $20\times20$ Si and Ge
monolayers, respectively; For comparison, the acoustic mismatch (AM)
model result of a single Si-Ge interface is also plotted.}
\end{figure}

Furthermore, when the interfaces are made of periodical superlattices
of Si and Ge, interesting phenomenon emerges.
Fig.~\ref{fig:modulation}$(a)$ shows a series of peaks appearing in the
transmission for the superlattice of alternating Si and Ge
monolayer. We find that this behavior can be understood as a resonance
effect. This kind of resonant tunneling transmission has been reported
in \cite{phyldgaard}. When the superlattice layer is composed of
several monolayers of Si and Ge, normal mode calculation shows that
resonant frequencies fall into different parts and band gaps are  
brought about by the repetition of superlattice structure. So it can
be seen that the total energy transmission is the acoustic mismatch
transmission modulated by the resonance.

We find that these resonance gaps will relate the thermal conductance
of superlattice with the thickness of the period layer.  We map the 1D
chain results to a 3D lattice using an average lattice constant $l_0 =
5.5\;$\AA\ for the cross section of the cell, and 8 chains per
conventional diamond-type cell. Our result of thermal conductance is
given as $G \approx 10^{9}\; {\rm Wm}^{-2}{\rm K}^{-1} $ at $200\,$K,
which agrees quantitatively with the measured effective values in
\cite{selee}. From Fig.~\ref{fig:exp}, we see that thermal conductance
decreases with the increase in the thickness of period $L$ at high
temperatures ($>100\,$K) due to the appearance of mini gaps in the
transmission, which agrees with the experimental result in
\cite{selee}.  We argue that in the several superlattice length scale
the thermal conductance is wave-natured ballistic transport because
the phonon mean free path is comparable to this length scale. But for
the samples of thickness of order $\mu$m \cite{selee}, the total
thermal resistance will be the diffusive summation of the microscopic
ballistic thermal resistance because wave coherence will be destroyed
after transporting across a few superlattice periods. So the effective
thermal conductance in \cite{selee} will exhibit microscopic ballistic
thermal transport behavior. Authors of Ref~\cite{superlattice}
interpret this kind of thermal conductance as ``partially coherent
heat conduction'' by introducing an imaginary wave vector. When the
thickness of the superlattice is larger than the phonon mean free
path, the conventional diffusive conductance comes into play and the
thermal conductivity will rise with increasing $L$ \cite{selee}.
These qualitative agreements between our calculations and the
experimental results suggest a ballistic heat transport as the dominating
mechanism within a few superlattice periods.  Another feature we find
from our calculation is that for lower temperatures ($<50\,$K), the
superlattice conductance becomes independent of the thickness and
tends to some universal value. We expect that some experiments will
verify this point in future.

In summary, we have derived a Landauer formula suitable for lattice
dynamic calculations.  The transmission coefficients are calculated by
a linear equation solver with proper boundary conditions.  We have
applied our methods to two types of junction structures, and give
significant results.

We thank Ming Tze Ong and Nan Zeng for critical readings of the
manuscript. 
This work is supported in part by a Faculty Research Grant of National
University of Singapore.

\end{document}